\documentclass{80SA}
\usepackage{times}

\title{High-Field de Haas-van Alphen Effect in non-centrosymmetric CeCoGe$_3$ and LaCoGe$_3$}

\author{Ilya Sheikin\thanks{E-mail address: ilya.sheikin@grenoble.cnrs.fr}, Pierre Rodiere$^{1}$, Rikio Settai$^{2}$, and Yoshichika \={O}nuki$^{2}$}
\inst{Grenoble High Magnetic Field Laboratory, CNRS, BP 166, 25
Av. des Martyrs, 38042 Grenoble, France \\
$^{1}$Institute N\'{e}el, CNRS-UJF, 25 Av. des Martyrs, 38042 Grenoble, France \\
$^{2}$Graduate School of Science, Osaka University, Toyonaka,
Osaka 560-0043, Japan}

\abst{We report on de Haas-van Alphen effect measurements in the non-centrosymmetric systems
CeCoGe$_3$ and LaCoGe$_3$ in magnetic field up to 28 Tesla. In both compounds, two new high
frequencies were observed in high fields. The frequencies were not detected in previous lower field
measurements. The frequencies do not originate from magnetic breakdown, and, therefore, are likely
to be intrinsic features of the compounds. In CeCoGe$_3$, the corresponding effective masses are
strongly enhanced, being of the order of 30 bare electron masses.}

\kword{CeCoGe$_3$, LaCoGe$_3$, non-centrosymmetric, de Haas-van Alphen effect}

\begin{document}
\maketitle


Following the discovery of superconductivity in the non-centrosymmetric heavy-fermion compound
CePt$_3$Si\cite{Bauer2004}, materials without inversion symmetry in their crystal structure have
attracted a lot of experimental and theoretical interest. The interest is mainly due to a
fascinating theoretical prediction\cite{Edelstein1989, Gor'kov2001} that superconductive pairing in
such systems requires an admixture of a spin-singlet with a spin-triplet state. Subsequent
observation of superconductivity in other non-centrosymmetric compounds
CeRhSi$_3$,\cite{Kimura2005} CeIrSi$_3$,\cite{Suginishi2006} CeCoGe$_3$,\cite{Settai2007,
Kawai2008} and CeIrGe$_3$\cite{Honda2010} has further stimulated the research efforts. Very unusual
superconducting properties were indeed observed in some systems, such as enormously high upper
critical field in CeIrSi$_3$\cite{Settai2008} and CeRhSi$_3$.\cite{Sugawara2010}

The absence of inversion symmetry in the crystal lattice of a metal brings about a strong
spin-orbit coupling, which in turn leads to the splitting of the electronic energy bands. The
Fermi-surface of such a metal is, therefore, also split into two surfaces characterized by
different chirality. This naturally results in the appearance of two distinct frequencies in the
spectra of de Haas-van Alphen (dHvA) oscillations. It was demonstrated
theoretically,\cite{Mineev2005} that the analysis of the oscillatory spectra in such materials
provides a direct measure of the strength of the spin-orbit coupling. It is thus very important to
obtain detailed and precise dHvA frequencies and their angular dependence in materials without
inversion center. In most non-centrosymmetric compounds this coupling is extremely strong being of
the order of 1000 K.

CeCoGe$_3$ and its non-4$f$ analog LaCoGe$_3$ crystallize in the tetragonal BaNiSn$_3$-type crystal
structure. CeCoGe$_3$ is a moderate heavy-fermion system with a Sommerfeld coefficient of the
specific heat $\gamma = 0.032\:$J/molK$^2$. It undergoes an untiferromagnetic transition at $T_{N1}
= 21\:$K, followed by two more transitions at $T_{N2} = 12\:$K and $T_{N3} =
8\:$K.\cite{Thamizhavel2005} Three consecutive metamagnetic transitions were observed for magnetic
field applied along the easy magnetic $c$-axis. The high pressure phase diagram of CeCoGe$_3$ is
quite complicated demonstrating six different phases.\cite{Knebel2009} Magnetic order vanishes
around 55 kbar, and superconductivity is observed in the range of 54\--–75 kbar.

The dHvA effect in CeCoGe$_3$ and LaCoGe$_3$ was previously investigated in magnet fields up to 17
Tesla.\cite{Thamizhavel2006} In both compounds, four fundamental frequencies were identified, all
of them split due to spin-orbit interaction. The frequencies themselves and their angular
dependencies are similar in both compounds and are in a rather good agreement with the results of
theoretical band structure calculations performed for LaCoGe$_3$. As compared to other
non-centrosymmetric compounds, the splitting of dHvA frequencies in CeCoGe$_3$ and LaCoGe$_3$ is
relatively small, implying a moderate spin-orbit coupling of about 100 K. In LaCoGe$_3$, all the
dHvA frequencies were doubled implying either a slightly unperfect quality of the sample or a field
dependence of the frequencies. The highest effective mass observed in CeCoGe$_3$ is 12 bare
electron masses corresponding to the $\beta$-branch. Finally, only one frequency originating from
the $\alpha$-branch representing the biggest Fermi surface was initially observed in
CeCoGe$_3$.\cite{Settai}


We present here the results of the dHvA effect measurements in non-centrosymmetric CeCoGe$_3$ and
LaCoGe$_3$ in magnetic fields up to 28 Tesla produced by a resistive magnet in GHMFL. The
measurements were performed on single crystals similar to those studied by Thamizhavel \textit{et
al}.\cite{Thamizhavel2006} The details of crystals preparation and characterization  are given
elsewhere.\cite{Thamizhavel2005} The measurements were performed using a torque cantilever
magnetometer. The magnetometer was mounted in a top-loading dilution refrigerator equipped with a
low-temperature rotation stage.


\begin{figure}
\begin{center}
\includegraphics{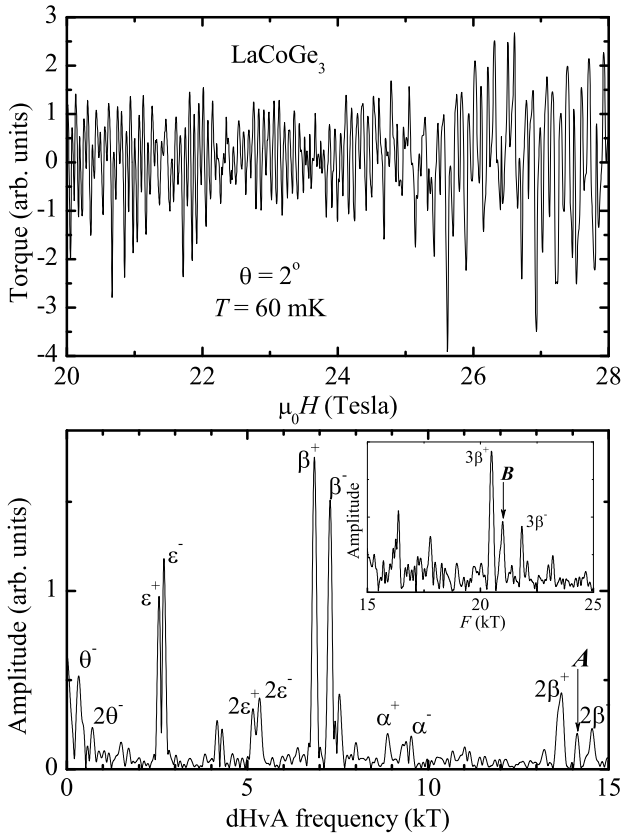}
\end{center}
\caption{High field (20\---28 T) dHvA oscillations (upper panel) and its Fourier spectrum (lower
panel) in LaCoGe$_3$ for magnetic field applied at 2$^\circ$ from the $c$-axis at 60 mK. The
frequencies that were observed in the previous lower-field measurements are denoted by Greek
letters, while the new frequencies observed at high field only are denoted $A$ and $B$.}
\label{LaCoGe3_FFT}
\end{figure}

Figure \ref{LaCoGe3_FFT} shows the oscillatory torque and the corresponding Fourier transform in
LaCoGe$_3$ in field from 20 to 28 Tesla. All the previous dHvA frequencies\cite{Thamizhavel2006}
can be clearly identified, and their values are very close. The most remarkable difference with
lower field measurements is the observation of two new fundamental frequencies denoted as $A$ and
$B$ in figure \ref{LaCoGe3_FFT}. The new frequencies, $F_A = 14.1$ kT and $F_B = 21$ kT, are
considerably higher than the highest frequency, $F_\alpha = 9.15$ kT, reported for lower fields.

\begin{figure}
\begin{center}
\includegraphics{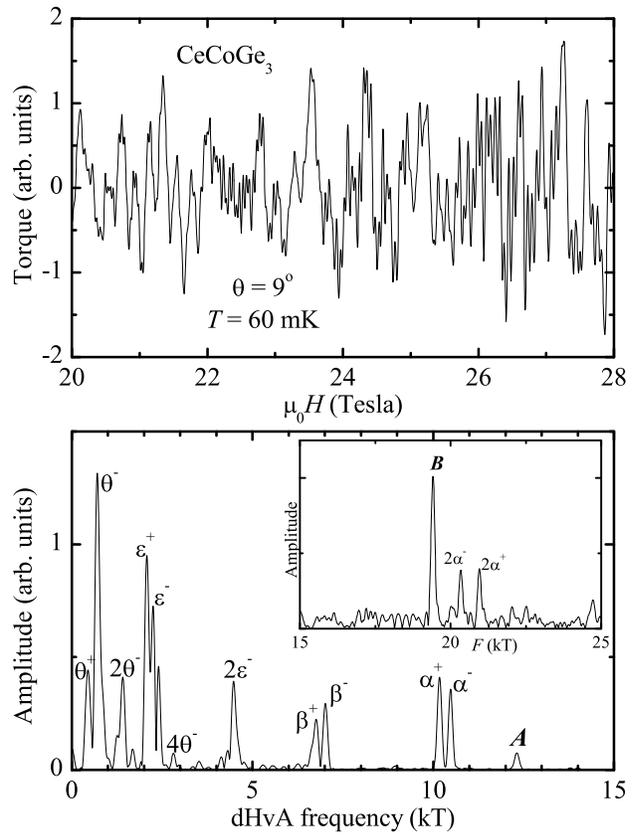}
\end{center}
\caption{dHvA oscillatory signal (upper panel) and its Fourier spectrum (lower panel) observed in
CeCoGe$_3$ with magnetic field (20\---28 T) applied at 9$^\circ$ from $[100]$ to $[110]$ at $T =
60$ mK. The frequencies denoted $\theta$, $\epsilon$, $\beta$ and $\alpha$ were observed in the
previous lower field measurements (with the exception of $\alpha^+$-branch). The frequencies
denoted $A$ and $B$ are observed only at high field and were not detected in the previous
measurements.} \label{CeCoGe3_FFT}
\end{figure}

The dHvA oscillations observed in CeCoGe$_3$ between 20 and 28 T are shown in the upper panel of
figure \ref{CeCoGe3_FFT}. Like in LaCoGe$_3$, all the fundamental frequencies previously observed
in CeCoGe$_3$ are still present at high field (lower panel of figure \ref{CeCoGe3_FFT}). In
addition, both components of the $\alpha$-branch are now clearly resolved. With magnetic field
applied at 9$^\circ$ from the crystallographic $c$-axis, the two frequencies are close to each
other, $F_{\alpha^+} = 10.17$ kT and $F_{\alpha^-} = 10.47$ kT. The most significant result,
however, is the presence of the new frequencies, $A$ and $B$, in the Fourier spectrum of
CeCoGe$_3$. The frequencies are somewhat lower than in LaCoGe$_3$ being $F_A = 12.31$ kT and $F_B =
19.42$ kT. Like in LaCoGe$_3$, both new frequencies are fundamental.

\begin{figure}
\begin{center}
\includegraphics{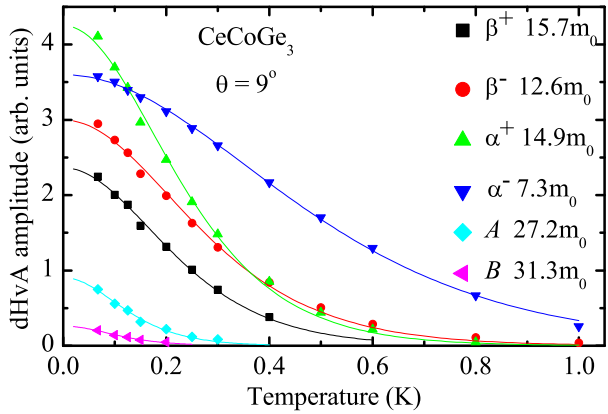}
\end{center}
\caption{(Color online) Temperature dependence of the dHvA amplitude is shown for $\alpha$ and
$\beta$-branches as well as the new frequencies $A$ and $B$ of CeCoGe$_3$ with magnetic field
applied at 9$^\circ$ from $c$ to $a$-axis. Lines are the fits to the temperature dependent part of
the Lifshits-Kosevich formula $A(T)  \propto  \frac{\alpha m^\ast T / B}{\sinh (\alpha m^\ast T
/B)}$, where $\alpha \approx 14.69\:$T/K. The effective masses, $m^\ast$, obtained from the fits
are also shown. $m_0$ is the bare electron mass.} \label{CeCoGe3_mass}
\end{figure}

Figure \ref{CeCoGe3_mass} shows the temperature dependence of the dHvA amplitudes of the new
frequencies $A$ and $B$ as well as $\alpha$ and $\beta$-branches in CeCoGe$_3$ for magnetic field
applied at 9$^\circ$ from the crystallographic $c$-axis, the same orientation as in figure
\ref{CeCoGe3_FFT}. These data allow one to determine effective masses by fitting the experimental
points to the temperature-dependent part of the Lifshits-Kosevich formula.\cite{Shoenberg1984} The
best fits to the formula along with the extracted effective masses are also shown in figure
\ref{CeCoGe3_mass}. It was previously reported\cite{Thamizhavel2006} that $\alpha$ and
$\beta$-branches possess the highest effective masses of 8 and 12 bare electron masses
respectively. These values are very close to the current results if only the $\alpha^-$ frequency
is considered and taking into account that only a single frequency from the $\alpha$-branch was
initially observed in previous measurements. Interestingly, the affective masses of the two
frequencies originating from the $\alpha$-branch differ by a factor of more than two, being 14.9
and 7.3 bare electron masses for $\alpha^+$ and $\alpha^-$ frequencies respectively. This is in
contrast with all the other branches where the effective masses of the two components are quite
close to each other. The most surprising result, however, is that the effective masses of the new
frequencies are strongly enhanced being 27 and 31 bare electron masses for $A$ and $B$
respectively. These values by far exceed the masses of the other previously observed frequencies.

\begin{figure}
\begin{center}
\includegraphics{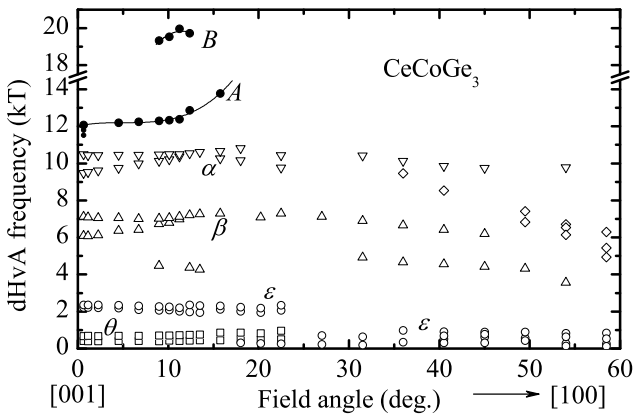}
\end{center}
\caption{Angular dependence of the dHvA frequencies in CeCoGe$_3$ is shown for the magnetic field
rotated from [001] towards [100] direction. Only fundamental frequencies are shown. The new
frequencies $A$ and $B$ are shown as closed symbols.} \label{CeCoGe3_angular}
\end{figure}

Figure \ref{CeCoGe3_angular} shows the angular dependence of the dHvA frequencies in CeCoGe$_3$ for
the magnetic field rotation from [001] towards [100] direction. The frequencies were determined
from the Fourier transform over the field range from 20 to 28 Tesla. The angular dependence of the
frequencies observed at lower fields is very similar to the previously reported
one.\cite{Thamizhavel2006} The highest of the new frequencies, $B$, exists only over a small
angular range at about 10$^\circ$ from the $c$-axis. The other new frequency, $A$, however, is
observed over an angular range of about 15$^\circ$ and is centered around the $c$-axis.


It is not perfectly clear whether the new dHvA frequencies $A$ and $B$ observed both in LaCoGe$_3$
and CeCoGe$_3$ emerge only at high field above 17 T or exist at lower fields as well, but can not
be detected due to experimental limitations. Very small amplitudes of the corresponding
oscillations as well as strongly enhanced effective masses found in CeCoGe$_3$ make the latter
scenario quite possible. In LaCoGe$_3$, however, while the amplitudes of $A$ and $B$ oscillations
are still relatively small, the effective masses must be small too. The new frequencies, if
present, should therefore be possible to observe at lower field. An appealing possibility is a
decrease of the Dingle temperature at high field. Since the oscillatory amplitude depends
exponentially on the Dingle temperature, the oscillations might be too small to detect at low
field, but become "visible" at higher field where the Dingle temperature is lower. Such scenario is
indeed realized in CeCoIn$_5$,\cite{Sheikin2006} where some of the dHvA frequencies are detected
only at high field above an abrupt change of the Dingle temperature. Nonetheless, if the new
frequencies exist, even if experimentally undetectable, already at low field, it would be difficult
to explain their absence in theoretical band structure calculations at least for LaCoGe$_3$, where
neither magnetic order nor 4$f$-electrons complicate the picture. Moreover, the Fermi-surface
cross-sections corresponding to both frequencies do not exceed the area of the Brillouin zone
perpendicular to the applied magnetic field. On the other hand, if the emergence of the new
frequencies is intrinsically related to high magnetic fields, this would easily account for their
absence in the band structure calculations performed for zero magnetic field. If this indeed is the
case, the new frequencies certainly do not originate from a magnetic breakdown as there are no two
fundamental frequencies that would sum up to yield any of them. Furthermore, the high effective
masses observed in CeCoGe$_3$ also rule out such a possibility. Thus, the new frequencies are
likely to be an intrinsic property of the compounds and seem to originate from a field-induced
modification of the Fermi-surface.

In conclusion, two new high dHvA frequencies have been observed both in LaCoGe$_3$ and CeCoGe$_3$
in high magnetic field up to 28 T. The frequencies are similar in both compounds. Neither of the
new frequencies was detected in the previous lower-field measurements or revealed by theoretical
band structure calculations. In CeCoGe$_3$, the frequencies correspond to strongly enhanced
effective masses implying that they represent thermodynamically important parts of the
Fermi-surface. While it is not certain if the frequencies appear in high magnetic field only or
simply experimentally undetectable at lower field, they certainly do not originate from magnetic
breakdown. They are therefore likely to be intrinsic for both compounds and are possibly due to a
field-induced modification of the Fermi-surface.


Part of this work has been supported by the EuroMagNET II under the EU contract number 228043.


\begin{thebibliography}{99}
\bibitem{Settai} The second dHvA frequency originating from the $\alpha$-brach was observed later
on by the same authors in subsequent experiments.
\end{thebibliography}
\end{document}